\documentclass[longauth]{aa}
\usepackage{txfonts}
\usepackage{graphicx}
\usepackage{longtable}

\begin{document}
\renewcommand{\topfraction}{0.85}
\renewcommand{\bottomfraction}{0.7}
\renewcommand{\textfraction}{0.15}
\renewcommand{\floatpagefraction}{0.66}

\title{H.E.S.S. upper limits for Kepler's supernova remnant}

\author{F. Aharonian\inst{1,13}
 \and A.G.~Akhperjanian \inst{2}
 \and U.~Barres de Almeida \inst{8} \thanks{supported by CAPES Foundation, Ministry of Education of Brazil}
 \and A.R.~Bazer-Bachi \inst{3}
 \and B.~Behera \inst{14}
 \and M.~Beilicke \inst{4}
 \and W.~Benbow \inst{1}
 \and D.~Berge \inst{1} \thanks{now at CERN, Geneva, Switzerland}
 \and K.~Bernl\"ohr \inst{1,5}
 \and C.~Boisson \inst{6}
% ...
}

\institute{
Max-Planck-Institut f\"ur Kernphysik, Heidelberg, Germany
\and
Yerevan Physics Institute, Armenia
\and
Centre d'Etude Spatiale des Rayonnements, CNRS/UPS, Toulouse, France
\and
Universit\"at Hamburg, Institut f\"ur Experimentalphysik, Germany
\and
Institut f\"ur Physik, Humboldt-Universit\"at zu Berlin, Germany
\and
LUTH, Observatoire de Paris, CNRS, Universit\'e Paris Diderot, Meudon, France
\and
IRFU/DSM/CEA, CE Saclay, Gif-sur-Yvette, France
\and
University of Durham, Department of Physics, U.K.
\and
Unit for Space Physics, North-West University, Potchefstroom, South Africa
\and
Laboratoire Leprince-Ringuet, Ecole Polytechnique, CNRS/IN2P3,
Palaiseau, France
\and 
Laboratoire d'Annecy-le-Vieux de Physique des Particules, CNRS/IN2P3,
Annecy-le-Vieux, France
\and
Astroparticule et Cosmologie (APC), CNRS, Universite Paris 7 Denis Diderot,
10, Paris, France
\thanks{UMR 7164 (CNRS, Universit\'e Paris VII, CEA, Observatoire de Paris)}
\and
Dublin Institute for Advanced Studies, Dublin, Ireland
\and
Landessternwarte, Universit\"at Heidelberg, K\"onigstuhl, Heidelberg, Germany
% ...
}

\abstract{}{Observations of Kepler's supernova remnant (G4.5+6.8) with the 
H.E.S.S. telescope array in 2004 and 2005 with a total live time of 13 h are 
presented.}{Stereoscopic imaging of Cherenkov radiation from extensive air 
showers is used
to reconstruct the energy and direction of the incident gamma rays.}
{No evidence for a very high energy (VHE: $>$100 GeV) gamma-ray signal from the 
direction of the remnant is found.
An upper limit (99\% confidence level) on the energy flux in the range 
$230 \, \mbox{GeV} - 12.8 \, \mbox{TeV}$ of $8.6 \times 10^{-13} \, \mbox{erg} 
\, \mbox{cm}^{-2} \, \mbox{s}^{-1}$
is obtained.}
{In the context of an existing theoretical model for the remnant, the lack of a
detectable gamma-ray flux implies a distance of at least 
$6.4 \, \mbox{kpc}$. A corresponding upper limit for the density
of the ambient matter of $0.7 \, \mbox{cm}^{-3}$ is derived.
With this distance limit, and assuming a spectral index $\Gamma = 2$, the total 
energy in accelerated protons 
is limited to $E_{p} < 8.6 \times 10^{49} \, \mbox{erg}$. 
In the synchrotron/inverse Compton framework, 
extrapolating the power law measured by RXTE between $10$ and $20 \,
\mbox{keV}$ down in energy, the predicted gamma-ray flux from 
inverse Compton scattering is below the measured upper limit for magnetic field
values greater than $52 \, \mu \mbox{G}$.}

\offprints{dominik.hauser@mpi-hd.mpg.de}
\maketitle

\section{Introduction} \label{intro}
It is widely believed that the bulk of the Galactic cosmic rays (CR) with
energies up to at least several $100 \, \mbox{TeV}$ originates from supernova
explosions (see for example \cite{1994A&A...287..959D}). This implies copious
amounts of very high energy (VHE: $>$100 GeV) nuclei and electrons in the
shells of supernova remnants (SNRs).  These particles can produce VHE gamma
rays in interactions of nucleonic cosmic rays with ambient matter, via inverse
Compton (IC) scattering of VHE electrons off ambient photons, as well as from
electron Bremsstrahlung on ambient matter. Therefore SNRs are promising targets
for observations of VHE gamma rays.  
\par
In October 1604 several astronomers,
among them Johannes Kepler, observed a ``new star'' which today is believed to
have been a bright supernova (SN) at the Galactic coordinates $l = 4.5
^{\circ}$ and $b = 6.8 ^{\circ}$.  The remnant of this supernova has since been
a target of observations covering the entire electromagnetic spectrum.  In the
radio regime, \cite{1988ApJ...330..254D} determined a mean
angular size of $\sim 200''$ and a mean expansion law $R
\propto t^{0.50}$, where $R$ is the radius and $t$ is the time. 
However, the expansion parameter $x=\dot{R} t/R$ varies considerably 
around the SNR shell, $0.35 < x < 0.65$, possibly indicating spatial 
inhomogenities in the circumstellar gas density. In a very recent paper by 
\cite{2008arXiv0803.4011V} these properties, and the general asymmetry of the 
remnant, have been basically confirmed through X-ray measurements. They also 
allowed the analysis of a high-velocity synchrotron filament in the eastern 
part of the remnant with $x=0.7$. 
\par
In addition, the distance $d$ to the SNR is
still under debate.  \cite{1999AJ....118..926R} report on an HI absorption
feature in VLA data and use the Galactic rotation model of
\cite{1989ApJ...342..272F} to calculate a lower limit $d > (4.8 \pm 1.4) \,
\mbox{kpc}$.  They also give an upper limit on the distance due to the lack of
absorption by an HI cloud at $6.4 \, \mbox{kpc}$. The authors remark that these
values involve uncertainties because of the proximity of Kepler's SNR to the
Galactic center. 
In contrast, \cite{2005AdSpR..35.1027S} and subsequently
\cite{2007ApJ...662..998B} have given a lower source distance of $d = 3.9 (+1.9
-0.9)$~kpc, from an absolute shock velocity $\sim 1660 \pm 120 \, \mbox{km} \,
\mbox{s}^{-1}$ derived from the H$\alpha$ emission line width of a
Balmer-dominated filament that is located in the northwestern region. The line
broadening, taken as an indication of the downstream thermal gas temperature,
was used to determine the shock velocity. We shall return to this question in
the discussion section. 
\par 
Finally, the type of the
supernova is not undisputed. From the reconstructed light curve
\cite{1943ApJ....97..119B} claimed that it was a type Ia SN, but
\cite{1985AJ.....90.2303D} argued that the light curve is also consistent with
a type II-L.  \cite{1989ApJ...347..925S} and \cite{1999PASJ...51..239K}
observed a relative overabundance of heavy elements that agrees with type Ia
nucleosynthesis models, while \cite{1994A&A...287..206D} saw more evidence that
Kepler's SNR is the remnant of a core-collapse SN.  Its position, $500 - 750 \,
\mbox{pc}$ above the Galactic plane, is more consistent with a type Ia than a
type II SN, as a SN of the latter type is expected to be confined to the region
of high gas density found in the plane. However, in the case of a core-collapse
event this might be explained through the model of a runaway star, as proposed
by \cite{1987ApJ...319..885B}.  More recently, theoretical modeling of the
detailed thermal line spectra obtained with \textit{XMM} 
(\cite{2004A&A...414..545C})
led \cite{2005ApJ...624..198B} to the conclusion that the X-ray spectrum is
best fit by a type Ia SN, a view also expressed by
\cite{2005ASPC..342..416B}. Most recently \cite{2007ApJ...668L.135R} reported
on deep \textit{Chandra} observations and argued from the high abundance of 
iron and the
very low abundance of oxygen that the progenitor of Kepler's SNR has been a
type Ia SN. Therefore it appears that the observational evidence is finally
converging on a type Ia event.  
\par 
In this paper observations of Kepler's SNR
with the H.E.S.S. telescope array are described.  An upper limit on the
integrated energy flux above $230 \, \mbox{GeV}$ is derived. Combining this
H.E.S.S. result with the theoretical predictions of \cite{2006A&A...452..217}
suggests a lower limit on the distance, close to the
upper limit given by \cite{1999AJ....118..926R}, if
Kepler's SN is a priori assumed to be of type Ia.
\section{H.E.S.S. Data and Analysis}
H.E.S.S. is an array of four imaging atmospheric Cherenkov telescopes situated
in the Khomas Highland of Namibia (\cite{2004NewAR..48..331H}).  Kepler's SNR
was observed with the entire telescope array between May 2004 and July 2005 for
a total observation time of 14 h.  The observations were made in \textit{wobble
mode}, where the tracking position is offset from the source center (RA
$17^{\mbox{\tiny h}}30^{\mbox{\tiny m}}42.12^{\mbox{\tiny s}}$, Dec
$-21^{\circ}28^{\prime}59.9^{\prime \prime}$ J2000.0). Offsets ranging from
$0.40^{\circ}$ to $0.85^{\circ}$ were used.  The data were taken at zenith
angles between $1^{\circ}$ and $49^{\circ}$, with a mean zenith angle of
$13^{\circ}$.  After applying the standard H.E.S.S. data-quality criteria a
total of $\sim$13~h live time were available for the analysis.  The analysis is
performed using the standard analysis techniques (\cite{2004APh....22..109A},
\cite{2005A&A...430..865A}).  
\par 
An event is counted as an ON-source event if
its direction is reconstructed within $0.112^{\circ}$ from the direction of the
source, given that Kepler's SNR is expected to be point-like for
H.E.S.S.\footnote{The value $0.112^{\circ}$ comes from a cut on the squared
angular distance of $0.0125 \, \mbox{deg}^{2}$ used in the standard
H.E.S.S. analysis.}  This is a reasonable assumption as the angular size of the
remnant in radio and X-rays wavelengths is $200'' (= 0.06^{\circ})$.  
\par 
As the
data were taken in wobble mode, the background estimation can be done using
OFF-source regions in the same field of view with the same size and offset
angle (angular distance to the pointing position) as the source region
(\cite{2007A&A...466.1219B}).  
\par 
A second independent analysis, used to
cross-check the results, is based on the three-dimensional modeling of the
Cherenkov light in the shower. (\cite{2006APh....25..195L}). The background
estimation for this second analysis was done similarly.  
\par
\begin{figure}[!t]

 \begin{minipage}[c]{4.3cm}
  \centering
  \includegraphics[width=4.3cm]{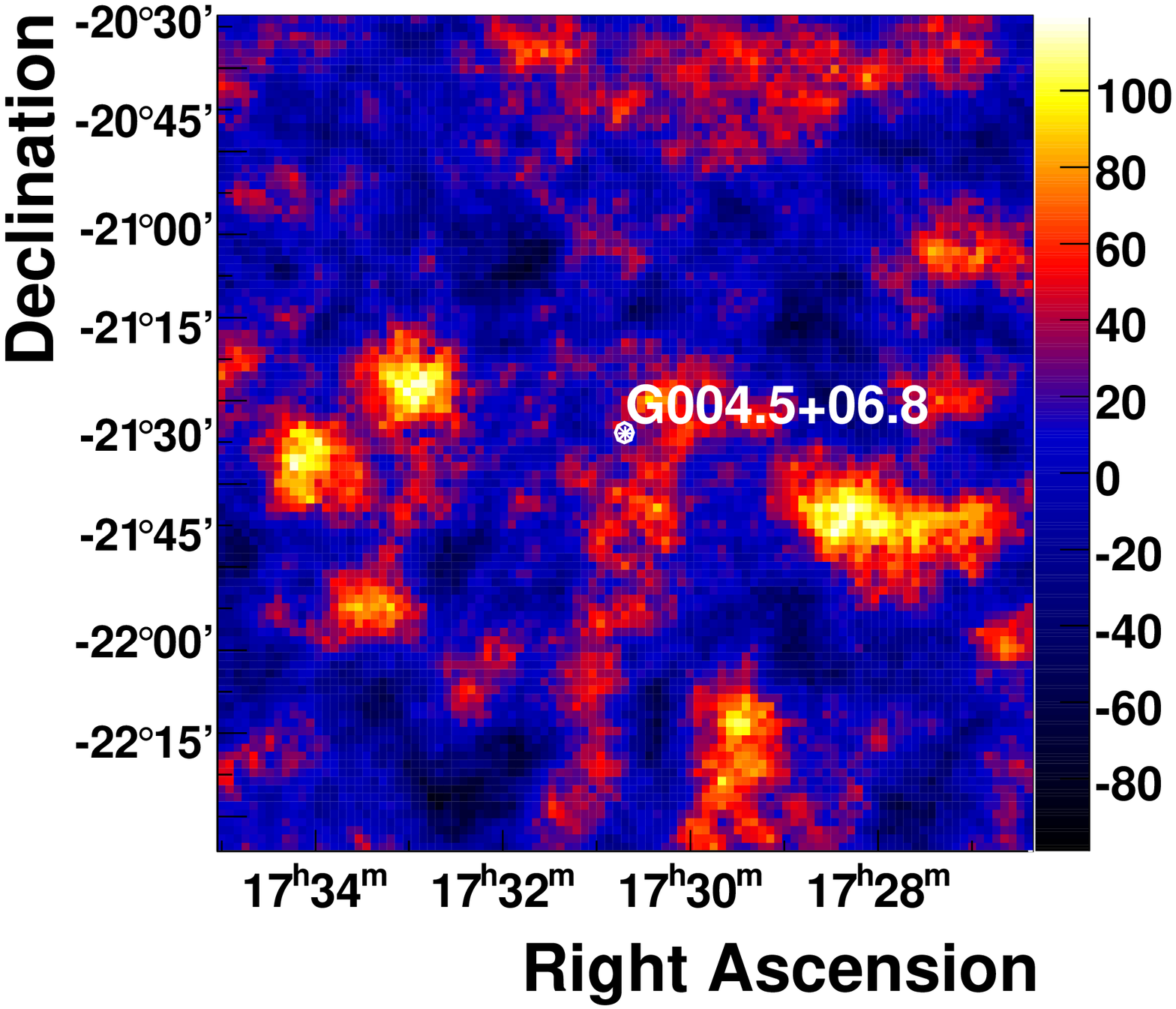}

 \end{minipage}
 \begin{minipage}[c]{4.3cm}
  \centering
  \includegraphics[width=4.3cm]{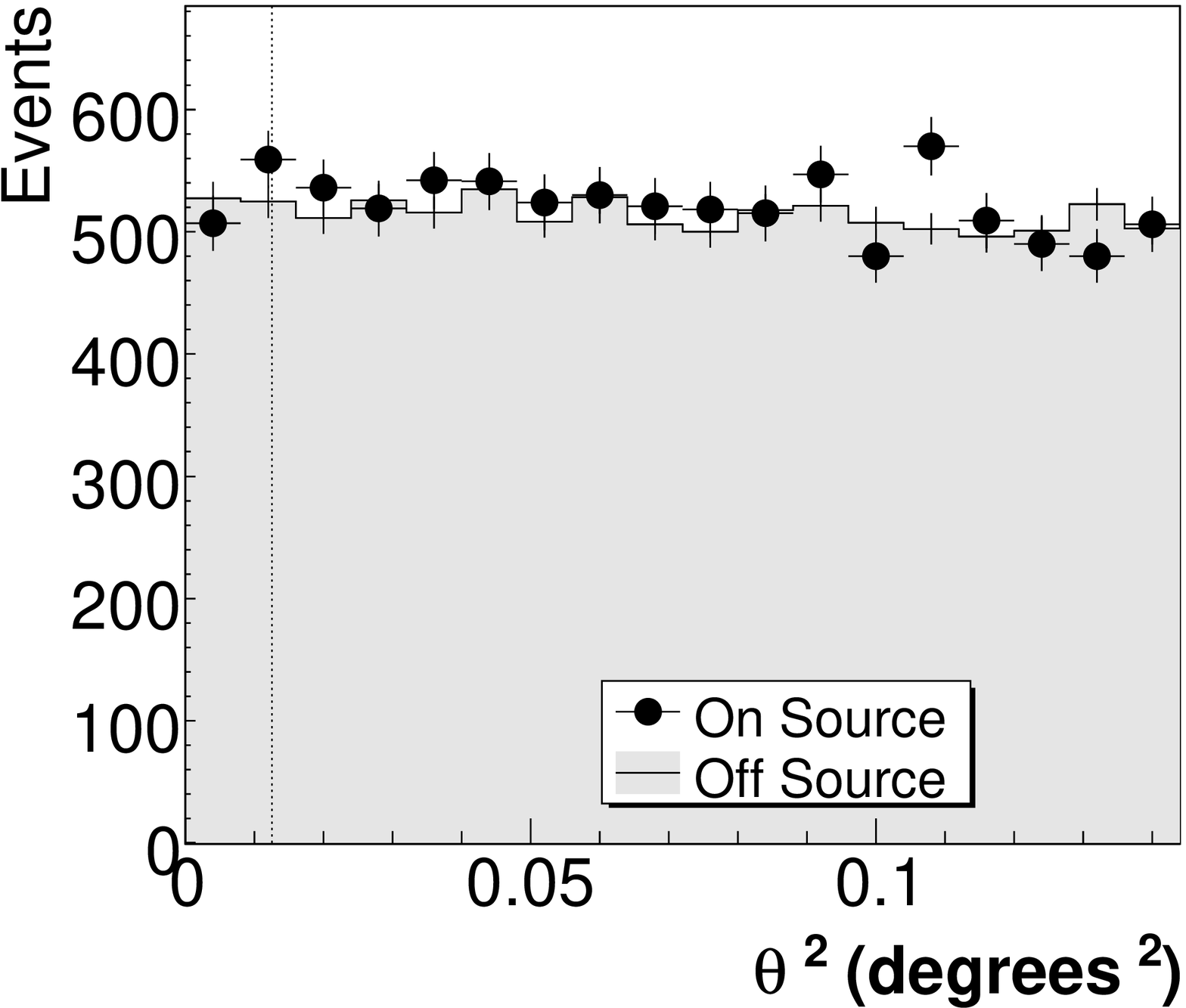}

  \end{minipage}

  \caption[]{\textbf{Left:} Sky map of excess events around the position of 
    Kepler's SNR with oversampling radius $0.112^{\circ}$; \textbf{Right:} 
    Distribution of the squared angular
    distance of gamma-ray-like events to the center of the remnant (ON) and
    the center of three control regions (OFF) with the same distance to the
    pointing position as the ON region. The vertical dotted line denotes the
    standard selection cut for point sources used by H.E.S.S.}
  \label{pics/excess_map_080}
\end{figure}

 
With the standard
analysis $827$ ON and $8855$ OFF events (with a normalization of $\alpha =
0.0911$) are measured, resulting in an excess of $20 \pm 30$ events. The total
significance of the excess from the direction of Kepler's SNR (calculated using
Equation 17 of \cite{1983ApJ...272..317L}) is $0.68$ standard deviations.
Fig.~\ref{pics/excess_map_080} shows in the left panel a sky map of excess
events around the position of Kepler's SNR and in the right panel the
distribution of the squared angular distance of observed gamma-ray candidates
from the center of the remnant in comparison to OFF data.
The angular distribution of the ON events is compatible with the distribution
of the OFF events. There is no evidence for a gamma-ray signal from Kepler's
SNR.  
\par
The approach of \cite{1998PhRvD..57.3873F} is used to calculate the upper 
limits on the integrated photon flux above $230 \, \mbox{GeV}$.  
At a confidence level of 99\%
an upper limit 
of 
$F(>230 \, \mbox{GeV}) < 9.3 \times 10 ^{-13} \, \mbox{cm}^{-2} \, 
\mbox{s}^{-1}$ 
for an assumed photon index of $\Gamma = 2.0$ is derived. At the same 
confidence 
level an upper limit on the energy flux of $F_{E}(230 \, 
\mbox{GeV} - 12.8 \, \mbox{TeV}) < 8.6 \times 10 ^{-13} \, \mbox{erg} \, 
\mbox{cm}^{-2} \, 
\mbox{s}^{-1}$ in the H.E.S.S. energy range for this data set 
($230 \, \mbox{GeV} - 12.8 \, \mbox{TeV}$) is derived. 
The assumed index of 2 requires an upper bound for the integration range to
avoid a divergent energy flux. 
\par
These values depend only weakly on the assumed photon index 
for reasonable values (i.e. $2.0 < \Gamma < 3.0$).
\section{Discussion}
\begin{figure}[!t]
  \centering \includegraphics[width=9cm]{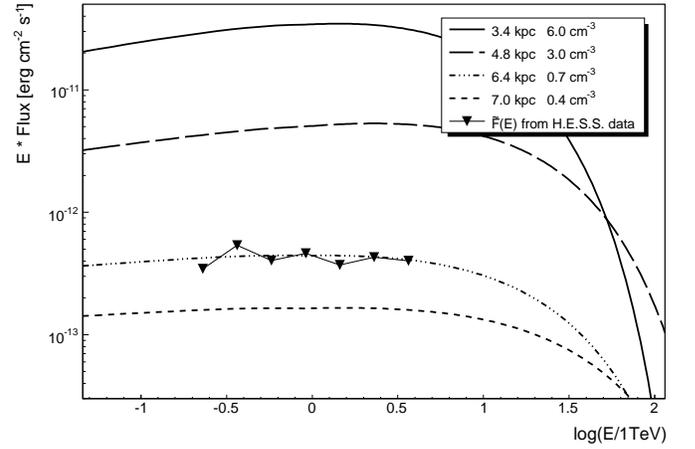}
  \caption{Comparison of the upper limits on $\tilde F (E)$ with predictions of
  BKV.}  
  \label{pics/upper_energy_limit_paper_070}
\end{figure}

To put the observed upper limit on the gamma-ray emission into perspective, the
above result is compared with theoretical
expectations. Such expectations have recently been formulated by
\cite{2006A&A...452..217} (BKV), using a non-linear kinetic theory of
cosmic-ray acceleration in SNRs. 
This model is based on a time-dependent, spherically symmetric
solution of the CR transport equation, coupled to the dynamics of the thermal
gas. 
The key assumption is that the explosion was 
a standard type Ia event in a
circumstellar medium at rest, representing an explosion energy $E_{\mbox{\tiny
SN}} \approx 10^{51} \, \mbox{erg}$ and an ejected mass of $1.4 \, M_{\odot}$. 
For a given distance the hydrogen density can then be derived from the
known angular expansion velocity and size of the remnant,
assumed to be given by the radio data of \cite{1988ApJ...330..254D} and
averaging these data over the azimuthal non-uniformities of the projected SNR
shell. The use of such an average value for the angular velocity of the shock
and the implied assumption of a uniform circumstellar density is a necessary
approximation within such a one-dimensional model which is meant to
describe the \textit{overall} physics of a point explosion. On the other hand, 
the systematic errors which these assumptions introduce are difficult to 
estimate, especially in the transition between sweep-up and adiabatic phase.
BKV obtained the spectrum and the spatial distribution of CR in the
remnant and the density of thermal gas. 
On this basis they then calculated the expected flux of non-thermal emission
(Fig.~\ref{pics/upper_energy_limit_paper_070}). To account for the
uncertainties in the distance estimate this was done for a distance range from
$3.4 - 7 \, \mbox{kpc}$.  The derived ambient density varies with the distance
assumed 
and the numerical results show that for a distance
$d$ as low as $4.8$~kpc the SNR has reached the Sedov
phase. Therefore the predicted integral hadronic gamma-ray flux
roughly decreases with distance $\propto E_{\mbox{\tiny SN}}^2
/d^{7}$, in agreement with the calculations shown in
Fig.~\ref{pics/upper_energy_limit_paper_070}. Approximating the emission
measure for free-free emission by $\mbox{EM} \sim N_{\mbox{\tiny H}}
M_{\mbox{\tiny sw}}$, where $M_{\mbox{\tiny sw}}$ denotes the swept-up
circumstellar mass, $\mbox{EM}$ scales in the same way with $E_{\mbox{\tiny
SN}}$ and $d$ as does the gamma-ray flux.  
\par 
To compare the given upper limit
with the model prediction, the quantity $\tilde F (>E) = E \cdot F(>E)$ is
determined. Here $F(>E)$ is the upper limit on the integrated Flux above the
energy $E$.  For $E = 230 \, \mbox{GeV}$ the value for $\tilde F$ is $\tilde
F(>230 \, \mbox{GeV}) = 3.4 \times 10^{-13}~\mbox{erg} \, \mbox{cm}^{-2} \,
\mbox{s}^{-1}$.  The resulting integrated upper limits are plotted in 
Fig.~\ref{pics/upper_energy_limit_paper_070} for several energies in the range 
$0.23 < (E/1 \, \mbox{TeV}) < 3.7$. Note that for these values no upper bound 
for the integration is needed as the quantity $F(>E)$ decreases with energy for
spectral indices greater than $\Gamma = 1.0$.
\par 
Within the context of the model of BKV, the H.E.S.S. upper limits rule out
distances smaller than $6.4~\mbox{kpc}$ for $E_{\mbox{\tiny SN}} \geq 10^{51}
\mbox{erg}$ and thus densities larger than $0.7 \, \mbox{cm} ^{-3}$,
and values of $\mbox{EM}$ in excess of $\geq 13 M_{\odot} \,
\mbox{cm}^{-3}$. 
The mean shock velocity is then $\approx 4000 \,
\mbox{km} \, \mbox{s}^{-1}$, and the SNR is just in transition from the
sweep-up phase to the Sedov phase.  
\par 
From SN explosion theory (see BKV, and references therein) 
a lower limit of $E_{\mbox{\tiny SN}} \sim 0.8
\times 10^{51} \, \mbox{erg}$ appears appropriate for type Ia SNe.  Considering
such a reduced explosion energy, the expected flux in gamma rays would be lower
and therefore the H.E.S.S.  upper limit would result in a reduced lower limit
on the distance of $d > 6.0 \, \mbox{kpc}$.  
\par
While in the BKV model the above value of $\mbox{EM}$ that corresponds
to the upper limit of H.E.S.S. agrees quite well with the overall number
recently derived by \cite{2007ApJ...662..998B} from their measurements, the
distances of 6.4 and 6.0 kpc differ 
significantly from the value adopted
by these authors, whose distance estimate is
within the errors smaller than 5.8 kpc (see Section~\ref{intro}). On the other 
hand,
\cite{2007ApJ...662..998B} derived their distance value from an optical
filament in the northwestern region which has the smallest expansion parameter
found in the radio and X-ray observations all around the remnant. 
It is also
interesting to note that the determination of the shock velocity from the
H$\alpha$ line broadening should involves a CR-modified shock, in contrast to
the assumption of \cite{2005AdSpR..35.1027S}. 
Efficient particle
acceleration in the SNR modifies the shock, whereby part of the gas compression
- but only a very small part of the gas heating - occurs in a smooth precursor,
in which the CR pressure gradient slows down the incoming gas flow. This is
followed by the so-called subshock (\cite{1981ApJ...248..344D})
where most of the gas heating occurs (\cite{1999ApJ...526..385B}).
The compression ratio $\sigma_\mathrm{s}$ of the subshock is smaller than the
overall shock compression ratio $\sigma$. Such a shock structure implies that
the shock velocity corresponding to the downstream thermal motions is the
subshock velocity $V_\mathrm{s}^\mathrm{sub} = \sigma_\mathrm{s} / \sigma \times
V_\mathrm{s}$, where $V_\mathrm{s}$ is the total shock velocity. In other words,
a higher overall shock velocity is required to achieve the same gas heating if
in addition CR are accelerated. Therefore the source distance derived from the
width of the H$\alpha$ line is $\sigma_\mathrm{s}/\sigma < 1$ times the true
source distance if derived without particle acceleration. This may imply a
substantial systematic error. In the BKV model for Kepler, in spherical
symmetry it is $\sigma_\mathrm{s}/\sigma \approx 0.4$ for an assumed distance 
of 4.8
kpc, and still equal to $0.6$ for $d=6.4$~kpc. Therefore the nominal distance
$d=4$~kpc adopted by \cite{2007ApJ...662..998B} is equivalent to $d = 6.6$~kpc,
if the northwestern region considered is indeed one where acceleration is
efficient. If particle acceleration is not efficient in this region, then the
correction factor is unity. Even a slight modification
$\sigma_\mathrm{s}/\sigma \approx 0.9$ of the shock makes the source distances
compatible.
\par
Independent of particle acceleration models one can use
the H.E.S.S. upper limit also to constrain the content of energetic particles 
in the remnant.  Using the limit on the flux in the range between $0.23 \,
\mbox{TeV}$ and $12.8 \, \mbox{TeV}$, an upper limit on the gamma-ray
luminosity $L_{\gamma, \mbox{\tiny max}} = 4 \pi F_{E} d^2$ can be estimated, where 
$F_{E}$ is the integrated energy flux upper limit. In this range then
$L_{\gamma, \mbox{\tiny max}} < 1.0 \times 10^{32} \cdot 
\left(d/\mbox{kpc}\right)^2 \, \mbox{erg s}^{-1}$ is derived. 
For power-law spectra, the $\delta$-function approximation 
\begin{equation}
\phi _{\pi} (E _{\pi}) \simeq \frac{c n}{K _{\pi}} \, \sigma
_{\mbox{\tiny pp}} \left( \frac{E _{\pi}}{K _{\pi}} \right) n _{\mbox{\tiny
p}} \left( \frac{E _{\pi}}{K
_{\pi}} \right)
\label{phi_e_pi} 
\end{equation}
can be used to relate the spectra of pions (or gamma rays) to those of the
primary protons (\cite{2000A&A...362...937}); here $\phi _{\pi}$ is the pion 
(or gamma-ray) production rate, $n$ the gas density, $n _{\mbox{\tiny p}}$ the 
number of protons and $K _{\pi}$ is the mean fraction of the kinetic energy of 
the proton transferred to the secondary $\pi ^{0}$-meson per collision. 
The rest mass of the proton is neglected. 
For spectral indices $\Gamma = 2-3$, $K _{\pi} = 0.17$ can be used
to approximate the pion spectrum (\cite{2000A&A...362...937}), a similar value
also applies for gamma-ray
spectra. At high energy, the proton-proton cross-section $\sigma _{\mbox{\tiny 
pp}}$ is only weakly energy dependent and can be approximated with
$\sigma_{\mbox{\tiny pp}} \simeq
40 \, \mbox{mb}$ (\cite{1990cup..book.....G}). The H.E.S.S. data probe proton
energies in the range from about $1$ to $100 \, \mbox{TeV}$; using equation
\ref{phi_e_pi} and assuming an index $\Gamma = 2.0$, a limit on the energy in
protons of $E _{\mbox{\tiny p}} < 4.9 \times 10 ^{47} \left( d/\mbox{kpc} \right)^2 \left( 
n/\mbox{cm}^{-3} \right)^{-1} \mbox{erg}$ can be derived. 
Extrapolating to the $1 \, \mbox{GeV}$ to $1 \, \mbox{PeV}$ range 
results in an upper limit 
on the total energy in accelerated protons of $1.5 \times 10^{48} \, 
\mbox{erg} \, \left( d/\mbox{kpc} \right)^{2} \left( n/\mbox{cm}^{-3}
\right)^{-1}$. 
Using $d = 6.4 \, \mbox{kpc}$ and $n = 0.7 \, \mbox{cm}^{-3}$ (dashed-3-dotted
line in Fig.~\ref{pics/upper_energy_limit_paper_070}) this results in 
$E _{\mbox{\tiny p}} < 8.6 \times
10^{49} \, \mbox{erg}$, i.e. $\sim 9\%$ of the assumed energy 
$E_{\mbox{\tiny SN}} = 10^{51} \, \mbox{erg}$.
This is of the order of what is expected for an average cosmic-ray
source in the form of a SNR. 
Assuming
$E_{\mbox{\tiny SN}} = 10^{51} \, \mbox{erg}$ and using the argument of
BKV that $d < 7 \, \mbox{kpc}$, in agreement with the observational argument of
\cite{1999AJ....118..926R}, the expected gamma-ray flux should not be lower
than the H.E.S.S. upper limit by more than a factor 2 as can be seen in 
Fig.~\ref{pics/upper_energy_limit_paper_070}.
\par
In another scenario the gamma-ray emission can be produced via IC 
scattering by VHE electrons off ambient photons mainly from the cosmic 
microwave background (CMB). 
The same electrons emit synchrotron X-ray radiation by being deflected
by magnetic fields in the SNR. The energy of the gamma-ray photons is
coupled to that of the X-ray photons
according to $\left(E_{\mbox{\tiny X}}/1 \, \mbox{keV} \right) \sim 0.07 \cdot
\left(E_{\gamma}/1 \,
\mbox{TeV} \right) \left(B/10 \, \mu \mbox{G} \right)$ in the case of the CMB
as target photon field for the IC scattering.
If the observed hard X-ray radiation (\cite{1999ICRC....3..480A})
, with a flux normalisation of $6.2 \times 10^{-5} \, \mbox{cm}^{-2} \, 
\mbox{s}^{-1} \, \mbox{keV} ^{-1}$ and a slope of $-3.0 \pm 0.2$, 
is synchrotron radiation (with the
corresponding energy flux\footnote{$f(E) = E ^{2} F(E)$} $f_{\mbox{\tiny X}}$)
the energy flux in gamma rays is 
given by $f_{\gamma} (E_{\gamma}) / f_{\mbox{\tiny X}}
(E_{\mbox{\tiny X}}) \sim \xi \, 0.1 (B/10 \, \mu \mbox{G}) ^{-2}$ 
(\cite{1997MNRAS.291..162A}). 
The factor $\xi$ takes into account possible 
differences in the source sizes in X-ray and gamma-ray wavelengths. 
We assume here $\xi \sim 1$. 
\par
In principle one could try to use the above relations to obtain a
lower limit on the magnetic field strength since the upper limit on
the flux in particular constrains any IC component.  For this purpose
the energy flux from X-ray synchrotron emission at an energy
corresponding to a given energy probed in VHE gamma rays has to be
known either from measurements or from detailed modeling.  The
interval in $E_{\mbox{\tiny X}}$ that corresponds to the observed
gamma-ray energy interval $0.23 < E_{\gamma}/1 \, \mbox{TeV} < 3.7$ is
$\sim (B/10 \, \mu \mbox{G}) \times (0.02 - 0.26) \, \mbox{keV}$,
whereas the energy interval in which the total non-thermal X-ray flux
is known is $10 - 20 \, \mbox{keV}$ (\cite{1999ICRC....3..480A}).  The
X-ray instrument PCA on board \textit{RXTE}, with which the underlying data
were obtained, has no imaging capabilities and therefore the measured
spectrum is the overall spectrum of the field of view of the
instrument (which is $1 ^{\circ}$).  Although it is expected that the
measured photon flux is indeed from Kepler's SNR because of its
position well above the plane, the X-ray flux has to be treated as an
upper limit.  Unfortunately there is no published analysis of the
non-thermal flux from \textit{Chandra} data covering the entire remnant. It is
also not possible to unambiguously disentangle the non-thermal and the
thermal contribution to the total spectrum measured by \textit{XMM-Newton} (G.
Cassam-Chena\"i, private communication).
\par
In the energy range around a few keV the extrapolation of the observed
hard X-ray flux to lower energies involves considerable
uncertainties. Nevertheless, in almost all scenarios the extrapolation
of the power-law spectrum measured between $10$ and $20 \, \mbox{keV}$
(with a spectral index of $\Gamma = 3.0$) should be an upper limit to
the X-ray to UV flux. With this extrapolation an upper limit on the
gamma-ray flux from IC scattering for a given magnetic field can be
calculated using the above formulas.
\par
For magnetic field values greater than $52 \, \mu \mbox{G}$ the 
resulting predicted upper limit on the IC flux would be less than the measured
upper limit of $F _{E} (3.7 \, \mbox{TeV}) < 2.91 \times 10 ^{-13} \,
\mbox{erg} \, \mbox{cm} ^{-2} \, \mbox{s} ^{-1}$. 
\par
From \textit{Chandra} measurements of thin X-ray filaments
(\cite{2005ApJ...621..793B}), whose thickness is interpreted as the 
synchrotron cooling length
of the radiating electrons, the actual field strength is $B \sim 300 \, \mu
\mbox {G}$, following the arguments of BKV and \cite{2006A&A...453..387P}. This
field implies an IC gamma-ray energy flux of $f _{\gamma} (E
_{\gamma}) < 1.4 \times 10 ^{-15} \, \mbox{erg} \, \mbox{cm} ^{-2} \, \mbox{s}
^{-1}$ which is two orders of magnitude below the measured upper limit.
 
\section{Conclusions}
Observations of Kepler's SNR with H.E.S.S. result in an upper limit
for the flux of VHE gamma rays from the SNR. In the context of an
existing theoretical model (BKV) for the remnant, and assuming an 
ejected mass of $1.4 \, M_{\odot}$ and an 
explosion energy of $10 ^{51} \,
\mbox{erg}$ in agreement with type Ia SN explosion models, the lack of
a detectable gamma ray flux implies a distance of at least $6.4 \,
\mbox{kpc}$, which is the same as the upper limit derived by
\cite{1999AJ....118..926R} from radio observations. 
Given that the gamma-ray flux effectively
scales with $E_{\mbox{\tiny SN}} ^{2}$, a significantly higher
explosion energy is excluded; a theoretically acceptable lower
explosion energy of $0.8 \times 10 ^{51} \, \mbox{erg}$ would lower
the distance limit to $6 \, \mbox{kpc}$.
\par
Assuming a purely hadronic scenario, a standard type Ia SN explosion, and using 
$6.4 \, \mbox{kpc}$ as a lower limit for the distance, the H.E.S.S. upper limit 
implies that the total energy in accelerated protons is 
less than $8.6 \times 10 ^{49} \, \mbox{erg}$.
\par
In a synchrotron/IC scenario no strong constraints on the magnetic field can be 
obtained.

\section*{Acknowledgments}
The support of the Namibian authorities and of the University of Namibia
in facilitating the construction and operation of H.E.S.S. is gratefully
acknowledged, as is the support by the German Ministry for Education and
Research (BMBF), the Max Planck Society, the French Ministry for Research,
the CNRS-IN2P3 and the Astroparticle Interdisciplinary Programme of the
CNRS, the U.K. Science and Technology Facilities Council (STFC),
the IPNP of the Charles University, the Polish Ministry of Science and 
Higher Education, the South African Department of
Science and Technology and National Research Foundation, and by the
University of Namibia. We appreciate the excellent work of the technical
support staff in Berlin, Durham, Hamburg, Heidelberg, Palaiseau, Paris,
Saclay, and in Namibia in the construction and operation of the
equipment.

\end{document}